\documentclass[
prl,
groupedaddress,
twocolumn
]{revtex4-1}

\usepackage{physics}
\usepackage{mathtools}
\usepackage[hyperindex]{hyperref}
\usepackage{amsmath}
\usepackage{xparse}
\usepackage{amssymb}
\usepackage{xcolor}
\usepackage[english]{babel}

\newcommand{\gkpkappa}{\kappa^{2}}

\renewcommand{\vec}[1]{\mathbf{#1}}

\newcommand{\logic}{\mathcal{L}}
\newcommand{\CZL}{\op{C}{}_\logic^Z}
\newcommand{\CZgen}[2]{\op{C}{}^Z_{#2}[#1]}

\newcommand{\qubitCS}[1]{\ket{\text{CS}_{#1}}}
\newcommand{\CVCS}[1]{\ket{\text{CVCS}_{#1}}}

\newcommand{\controlled}[1]{\op{C}_{#1}}
\renewcommand{\op}[1]{{#1}}

\def \brasubspacing{\hspace{-2pt}}
\def \ketsubspacing{\hspace{-1.5pt}}

\newcommand{\lsub}[2]{{\protect\vphantom{#1}}_{#2} \brasubspacing {#1}}

\DeclareDocumentCommand\bra{ s m O{} t\ket s g } %
{
    \IfBooleanTF{#4}
        { %
            \IfBooleanTF{#1}
            {%
                \IfNoValueTF{#6}
                    {\braket*{#2}{}\IfBooleanTF{#5}{*}{}}
                    {\braket*{#2}{#6}}
            }
            {
                \IfBooleanTF{#5}
                { %
                    \IfNoValueTF{#6}
                        {\braket{#2}{}*}
                        {\braket*{#2}{#6}}
                }
                {\braket{#2}{\IfNoValueTF{#6}{}{#6}}}
            }
        }
        { %
            {\lsub{\vphantom{#2}\left\langle
            \IfBooleanTF{#1}{\smash}{}{#2}\right\rvert}{#3}}
            \IfBooleanTF{#5}{*}{}
            \IfNoValueTF{#6}{}{#6}
        }
}

\DeclareDocumentCommand\ket{ s m O{} t_} %
{
    \IfBooleanTF{#4}
        {\vphantom{#2}\left\lvert\IfBooleanTF{#1}{\smash}{}{#2}
            \right\rangle_ }
        {\vphantom{#2}\left\lvert\IfBooleanTF{#1}{\smash}{}{#2}
            \right\rangle_{\ketsubspacing #3}}
}

\DeclareDocumentCommand\innerproduct{ s m g O{} o t_ } %
{
    \lsub{
        \left\langle
        \IfBooleanTF{#1}{\smash}{}{#2}
        \middle\vert
        \IfBooleanTF{#1}
            {\smash{\IfNoValueTF{#3}{#2}{#3}}}
            {\IfNoValueTF{#3}{#2}{#3}}
        \right\rangle
        }{#4}
        \IfNoValueTF{#5}
             {\IfBooleanTF{#6}{}{}}
             {_{\ketsubspacing #5}}
}

\DeclareDocumentCommand\outerproduct{ s m g O{} O{} } %
{
    \IfBooleanTF{#1}
        { %
            \IfNoValueTF{#3}
            {\ket*{#2}[#4]\!\bra*{#2}[#5]}
            {\ket*{#2}[#4]\!\bra*{#3}[#5]}
        }
        { %
            \IfNoValueTF{#3}
            {\ket{#2}[#4]\!\bra{#2}[#5]}
            {\ket{#2}[#4]\!\bra{#3}[#5]}
        }
}

\newcommand{\pket}[1]{\ket{#1}[p]}
\newcommand{\qket}[1]{\ket{#1}[q]}

\newcommand{\gpket}[1]{\ket{#1}[p,G]}
\newcommand{\gqket}[1]{\ket{#1}[q,G]}
\newcommand{\gket}[1]{\ket{#1}[G]}
\newcommand{\gbra}[1]{\bra{#1}[G]}

\newcommand{\mG}{m_G} %
\newcommand{\mGop}{\op{m}_G} %
\newcommand{\uG}{u_G}
\newcommand{\uGop}{\op{u}_G}

\renewcommand{\op}[1]{\hat{#1}}

\newcommand{\q}{\ope{q}}

\newcommand{\p}{\ope{p}}

\newcommand{\ufrac}{\op{u}}

\renewcommand{\u}{\ufrac}

\newcommand{\uint}{\op{m}}

\newcommand{\m}{\uint}

\DeclareFontFamily{U}{wncy}{}
\DeclareFontShape{U}{wncy}{m}{n}{<->wncyr10}{}
\DeclareSymbolFont{mcy}{U}{wncy}{m}{n}
\DeclareMathSymbol{\Sh}{\mathord}{mcy}{"58}
\newcommand{\reals}[0]{\mathbb{R}}
\newcommand{\complex}[0]{\mathbb{C}}

\newcommand{\integers}[0]{\mathbb{Z}}

\newcommand{\CZ}[0]{\controlled{Z}}

\newcommand{\X}{\op{X}}
\newcommand{\Z}{\op{Z}}

\DeclareDocumentCommand\fracpart{ s m g }
{
    \IfBooleanTF{#1}{\Bqty*}{\Bqty}{{#2}}\IfNoValueTF{#3}{}{_{#3}}}
\DeclareDocumentCommand\closestint{ s m g }
{
    \IfBooleanTF{#1}{\smash}{\left} \lfloor {#2}
    \IfBooleanTF{#1}{\smash}{\right} \rceil\IfNoValueTF{#3}{}{_{#3}}
}

\newcommand{\ope}{\op}

\DeclareDocumentCommand\dyad{}{\outerproduct}
\DeclareDocumentCommand\ketbra{}{\dyad}
\DeclareDocumentCommand\braket{}{\innerproduct}

\newcommand{\half}{\frac{1}{2}}

\newcommand{\mat}[1]{\mathbf{#1}}

\newcommand{\ketsub}[2]{\ket{#1}[#2]}

\begin{document}
\title{
Modular Bosonic Subsystem Codes
}

\author{Giacomo Pantaleoni}
\email{gpantaleoni@null.net}
\affiliation{Centre for Quantum Computation \& Communication Technology, School of Science, RMIT University, Melbourne, VIC 3000, Australia}
\author{Ben Q. Baragiola}
\affiliation{Centre for Quantum Computation \& Communication Technology, School of Science, RMIT University, Melbourne, VIC 3000, Australia}
\author{Nicolas C. Menicucci}
\affiliation{Centre for Quantum Computation \& Communication Technology, School of Science, RMIT University, Melbourne, VIC 3000, Australia}

\date{\today}

\begin{abstract}
 We introduce a framework to decompose a bosonic mode into two virtual subsystems---a logical qubit and a gauge mode.  This framework allows the entire toolkit of qubit-based quantum information to be applied in the continuous-variable setting.  We give a detailed example based on a modular decomposition of the position basis and apply it in two situations. First, we decompose Gottesman-Kitaev-Preskill grid states and find that the encoded logical state can be damaged due to entanglement with the gauge mode. Second, we identify and disentangle qubit cluster states hidden inside of Gaussian continuous-variable cluster states.
\end{abstract}

\maketitle

{\bf Introduction.---}Continuous-variable (CV) quantum computing is experiencing considerable theoretical \cite{Fukui:2017aa, Sabapathy:2018aa, Douce:2019aa,Vuillot:2019aa,Baragiola:2019aa,Noh:2019aa} and experimental \cite{Chen:2014jx,Ofek2016,Hu:2019aa, Fluhmann:2019aa, Gao:2019aa, Warit-Asavanant:2019aa, Larsen:2019aa} development due to the promise of substantial scalability.  Practical quantum computing, however, usually employs finite-dimensional Hilbert spaces, because error correction and fault tolerance require digital quantum information \cite{Aharonov:2008aa, Campbell:2017aa}.

Bosonic codes \cite{Lau:2016aa, Albert2018a} restore the notion of a qubit in CV quantum computing by identifying a discrete-variable Hilbert space ($\complex^{2}$ for qubits) within one or more bosonic modes.  The standard approach selects two wavefunctions that define the logical subspace, with the remaining ``wilderness space'' serving as as resource for error detection. This corresponds to a \emph{subspace} decomposition of the CV Hilbert space, $\mathcal{H}_\text{CV} = \mathbb{C}^2 \oplus \mathcal{H}_\text{wild}$.  A notable example is the Gottesman-Kitaev-Preskill (GKP) encoding \cite{gkp2001}, which can be combined with measurement-based quantum computing using CV cluster states to achieve fault tolerance \cite{Menicucci:2014cx}.

Bosonic subspace encodings suffer from several problems. In contrast to qubit-based subspace codes, a bosonic-code subspace is vanishingly small compared to the full CV Hilbert space.
In the case of GKP, the codewords themselves are unphysical, and approximate states can have little overlap yet represent the same logical information~\cite{TzitBourMeni20, Matsuura:2019aa, Wan:2019aa}.
A formal way to account for this fact has been lacking.  A step towards a solution was given in Refs \cite{Vernaz-Gris:2014aa, ketterer2016}, which presented a continuous direct-sum decomposition of the CV Hilbert space using a set of eigenstates for modular position and momentum.

In this Letter, we describe  a framework  to encode a qubit in an entirely different way: by identifying a discrete, logical \emph{subsystem} using a modular decomposition of the position basis. The resulting modular bosonic subsystem code gives a precise description both of a logical qubit hiding inside the CV Hilbert space and also of the complementary gauge mode. In contrast to bosonic subspace codes \cite{Cochrane:1999aa, Michael:2016aa, Albert2018a}, a bosonic subsystem code endows every CV state with logical-qubit information. This is more practical, as one is not required to construct operations (measurements and gates) that select only a particular subspace. Once a trace over the gauge mode is performed, the CV nature of the mode can be forgotten, and one can work entirely at the qubit level.

{\bf Bosonic subsystem  decomposition.}---Consider the division of a finite-dimensional Hilbert space $\mathcal{H} = \mathbb{C}^{dN}$ into $d$ orthogonal subspaces, $\mathcal{H} = \bigoplus_{i=1}^d \mathcal{H}_i$.  When $\dim( \mathcal{H}_i) = N$ for all $i$, the subspaces are isomorphic, and an alternate decomposition of the same Hilbert space is given by $\mathcal{H} = \mathbb{C}^d \otimes \mathbb{C}^N$. Interpreted as two virtual subsystems---a dimension-$d$ qudit and a dimension-$N$ gauge subsystem---a tensor-product basis can be constructed \cite{Raynal:2010aa, Albert:2019aa}. This works equally well for infinite-dimensional Hilbert spaces, with the additional feature that the gauge-subsystem Hilbert space can be isomorphic to $\mathcal{H}$.

We exploit this idea to encode a qubit (or qudit) into a bosonic mode. Since the Hilbert space of a mode, $\mathcal{H}_{\rm CV}$, is infinite dimensional, it can be decomposed into a qubit $(d=2)$ and another CV Hilbert space that is distinct from---but isomorphic to---the original mode, $\mathcal{H}_{\rm CV} = \mathbb{C}^2 \otimes \mathcal{H}_{\rm CV}$.  We refer to the division of a mode into two subsystems as a \emph{subsystem decomposition}.

Subsystem decompositions are not unique (see Conclusion).  Here, we construct an illustrative example based on applying modular arithmetic to the position basis. This choice allows us to identify a logical-subsystem qubit that describes and connects two seemingly disparate objects in CV quantum computation: the GKP encoding and continuous-variable cluster states.

\begin{figure}[t]
    \includegraphics[width=0.45\textwidth]{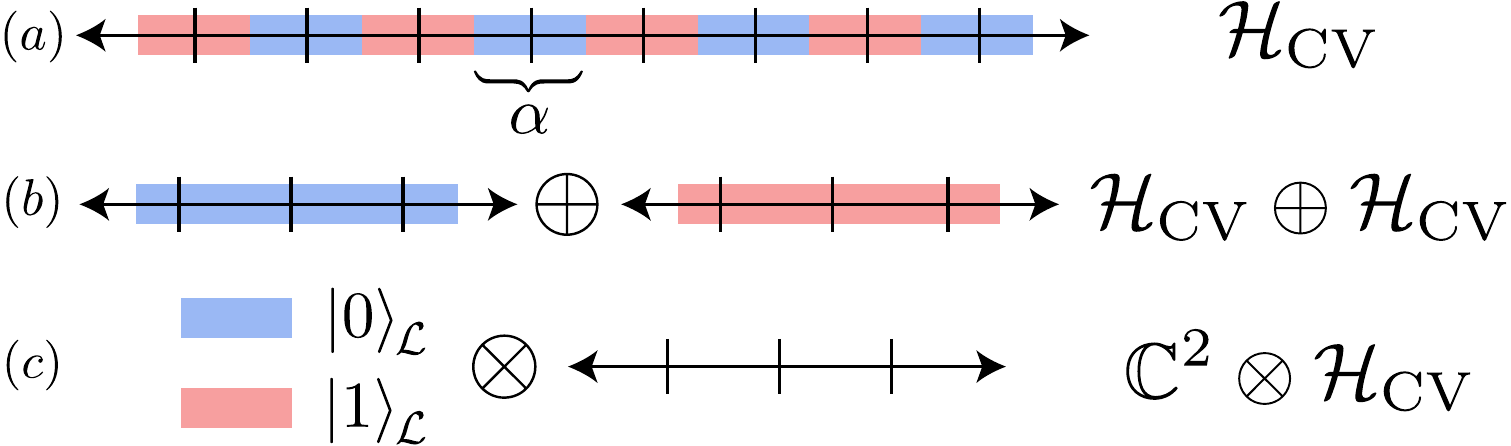} %
    \caption{Subsystem decomposition of a single mode.
(a) Modular decomposition of the position spectrum (the real line) into bins of size $\alpha$. Each bin is centered at $m \alpha$, with $m$ labeling the bin and $u\in[-\alpha/2,\alpha/2)$ giving the fractional remainder. Blue/red indicate even/odd values of $m$.
(b) The even- and odd-$m$ subspaces are each isomorphic to the original Hilbert space.
(c) Decomposition into a tensor product of a qubit subsystem and a gauge mode, where the parity of $m$ defines a logical qubit.
    }\label{fig:subsystem}
\end{figure}

{\bf Bosonic subsystem codes based on modular position.}---A real number $s \in \mathbb{R}$ can be split into its integer and fractional parts with respect to a ``bin size" $\alpha \in \reals$ as $s = \closestint{s}{\alpha} + \fracpart{s}{\alpha}$, where $\closestint{s}{\alpha} \coloneqq \alpha \lfloor \frac{s}{\alpha} + \frac{1}{2} \rfloor$ indicates closest (centered) integer multiple of $\alpha$ and $\fracpart{s}{\alpha} \coloneqq s - \closestint{s}{\alpha}$ is the (centered) remainder, see Fig.~\ref{fig:subsystem}(a). We call $\closestint{\cdot}{\alpha}$ the integer part and $\fracpart{\cdot}{\alpha}$ the fractional part of the object inside.  Using the spectral theorem, any quadrature operator can be decomposed in this way, such as the position operator $\q = \frac{1}{\sqrt{2}}(\op{a} + \op{a}^\dagger)$. Specifically, $\q = \alpha \m + \u,$ where $\alpha \m \coloneqq \closestint{\q}{\alpha}$ is the integer part of~$\op q$, and $\u \coloneqq \fracpart{\q}{\alpha}$ is its fractional part. Decomposing position eigenstates, $\op{q} \qket{s} = s \qket{s}$, we construct simultaneous eigenvectors of $\m$ and $\u$,  $\qket{s} = \qket{\alpha m  + u} \eqqcolon \ket{m, u}$. We call this the \emph{partitioned-position basis} since it partitions the position value $s=\alpha m + u$ into a \emph{bin number}~$m \in \mathbb{Z}$ and a \emph{modular position}~$u \in [ -\alpha/2,\alpha/2 )$~\cite{Aharonov1969}.  This basis is convenient for representing states with discrete translation symmetry---notably the codewords of the GKP encoding of a qubit in an oscillator~\footnote{ A Fourier series over $m$ gives the \emph{kq}-representation \cite{zak1967finite,Englert:2006aa} useful for analysis of GKP codes \cite{Glancy2006, Terhal2016, ketterer2016}}.

We identify a logical-qubit subsystem using the parity~$P(\cdot)$ of the bin-number operator $\m$, see Fig. \ref{fig:subsystem}.
Decomposing the position operator (and its eigenstates), $ \q = \alpha \op{\ell} +  2\alpha \mGop + \uGop$ with $\op{\ell} := P\pqty{\op{m}}$, $\uGop \coloneqq \u$, and $\mGop \coloneqq \tfrac 1 2 (\op m - \op \ell)$, gives the basis states,
	\begin{equation} \label{eq:subsystem_basisstate}
		\qket{s} = \ket{m,u} =  \ket{\ell}[\logic] \otimes \gket{\mG, \uG}
		 \, .
	\end{equation}
The logical qubit (denoted by subscript~$\logic$) emerges from the $m$-parity states $\ket{0}[\logic]$ and $\ket{1}[\logic]$.  The $m$-parity operator $\op{\ell}$ acts as a projector: $\op{\ell} = \ket{1}_ \logic\! \bra{1}[\logic] = \half \pqty*{ \op{I}_\logic - \Z_\logic}$, where $\op{I}_\logic$ and $ \Z_\logic$ are logical identity and Pauli-$Z$ operators, respectively.  The states $\gket{\mG, \uG}$, with $\mG \in \mathbb{Z}$ and $\uG \in [ -\alpha/2,\alpha/2 )$ constitute the partitioned-position basis for the gauge mode, denoted by subscript $G$.  (One can also define a gauge position basis $\ketsub {s_G}{q,G}$, where $s_G = \alpha \mG + \uG$.)

Together, the states $\ket{\ell}_\logic$ and $\gket{\mG, \uG}$ form a \emph{subsystem basis} for the joint Hilbert space of the logical qubit and the gauge mode, $\op{I}_\text{CV} = \op{I}_\logic \otimes \op{I}_G = \sum_\ell \ket{\ell}_ \logic\! \bra{\ell}[\logic] \otimes\sum_{\mG} \int d\uG \gket{\mG, \uG}\! \gbra{\mG, \uG} $.  Quantum operations can be decomposed in
this basis to interpret their actions on the two subsystems.  Operations affecting only the gauge mode do not disturb the encoded information, although physical CV operations typically affect both subsystems and may entangle them.

In a subsystem basis, any pure CV state $\ket{\Psi}$ has a Schmidt decomposition
	\begin{equation} \label{eq:decomposedpurestate}
		\ket{\Psi} = \sqrt {p_a} \ket{\psi_a}[\logic] \otimes \gket{\phi_a} + \sqrt{p_b} \ket{\psi_b}[\logic] \otimes \gket{\phi_b} \, ,
	\end{equation}
with Schmidt coefficients~$\sqrt{p_j} \geq 0$. When only one of these coefficients is nonzero, Eq.~\eqref{eq:decomposedpurestate} is a tensor-product state  encoding a pure logical qubit;  see examples in Table~\ref{tab:table1}.  More generally, the logical qubit and gauge mode are entangled.   Further, the CV state $\op{\rho}$ can be mixed.  The logical state $\op{\rho}_\logic$ is isolated by tracing over the gauge mode:
	\begin{equation} \label{eq:logicalstate}
		\op{\rho}_\logic =
		\Tr_G [ \op{\rho} ] = \sum_{\mG \in \mathbb{Z}} \int^{\alpha/2}_{-\alpha/2} d\uG \,
	\gbra{\mG,\uG}
	\op{\rho}
	\gket{\mG, \uG}
        \ ,
	\end{equation}
 with
 any entanglement between subsystems decohering the logical state.  In analogy to averaging over error subspaces in a qubit subsystem code \cite{RahnDoheMabu02}, the gauge trace isolates the qubit state within \emph{any} CV state (for a given subsystem decomposition).  This provides an advantage over bosonic subspace codes, which can fail to capture encoded information in a CV state, for instance when the state (typically a damaged codeword) has little or no support in the code's subspace.

A measure to compare two encoded CV states $\op{\rho}, \hat{\sigma}$ is given by the \emph{logical fidelity},
	\begin{equation} \label{eq:logicalfidelity}
            \mathcal{F}_\logic (\op{\rho}_\logic, \hat{\sigma}_\logic) \coloneqq \Big( \Tr \Big[ \sqrt{ \sqrt{\hat{\sigma}_\logic} \op{\rho}_\logic \sqrt{\hat{\sigma}_\logic} } \, \Big] \Big)^2 \, ,
	\end{equation}
where $\op{\rho}_\logic, \hat{\sigma}_\logic$ are obtained from the CV states via the gauge trace.  This logical fidelity was used in Ref. \cite{TzitBourMeni20} to show that two orthogonal CV states can encode the same logical state with $\mathcal{F}_\logic \rightarrow 1 $.

%
\begin{table}[t]
  \begin{center}
    \begin{tabular}{c|c|c} 
       ~CV state~ &  ~Partitioned-position basis~ & ~Subsystem basis~ \\
            \hline
      $\qket{0}$ & $\ket{0,0}$ & $ \ket{0}[\logic] \otimes \gqket{0}$
      \\
      $\pket{0}$ & $\sum_m \int du \ket{m,u}$ & $ \ket{+}[\logic] \otimes \gpket{0}$
      \\
	$\ket{\ell_\text{GKP}}$ & $\sum_{m} \ket{2m+\ell,0}$ & $ \ket{\ell}[\logic] \otimes  \gket{+_\text{GKP}}$
	\\
    \end{tabular}
        \caption{Representations of various (unnormalized) CV states, with sums taken over $\mathbb{Z}$ and integrals over the interval $[-\alpha/2,\alpha/2)$.
        } \label{tab:table1}
  \end{center}
\end{table}
%
{\bf Subsystem description of the GKP encoding.---} The subsystem decomposition in Eq.~\eqref{eq:subsystem_basisstate}  gives new a perspective on bosonic error-correcting codes.  We focus on the GKP encoding \cite{gkp2001}, notable for its favorable error-correcting properties and all-Gaussian gate set \cite{Terhal2016, Noh:2018aa, Fluhmann:2019aa, Baragiola:2019aa}.    The GKP codewords, $\ket{\ell_\text{GKP}} = \sum_{m \in \mathbb{Z}} \ket{(2m +\ell)\alpha}_q$ with $\ell \in \{0,1\}$, have position-space wavefunctions that are Dirac combs with spacing $2\alpha$.   Due to their periodic structure, each of these states is compactly represented in the partitioned-position basis with spacing $\alpha$; see Table \ref{tab:table1}. In the subsystem basis, Eq.~\eqref{eq:subsystem_basisstate}, GKP states are product states between the logical and gauge subsystems: $\ket{\psi_{\text{GKP}}} = \ket{\psi}[\logic] \otimes \gket{+_{\text{GKP}}}$. In fact, \emph{any} $2\alpha$-periodic state is a product state in this basis---the logical and gauge subsystems are unentangled. An example is the momentum eigenstate, $\pket{0} = (2\pi)^{-1/2}\int ds \, \qket{s} = \ket{+}[\logic] \otimes \gpket{0}$, which encodes the same logical information as $\ket{+_\text{GKP}}$ but with a different gauge-mode state. That is, both $\ket{+_\text{GKP}}$ and $\pket{0}$ encode the same outcome probabilities for the computational-basis measurements laid out in the original GKP formulation: binned homodyne detection~\cite{gkp2001}.  Differences arise when the states are transformed under logical operations or suffer errors---a state whose gauge mode is near $\ket{+_{\text{GKP}}}_G$ will be more resilient to errors (see below).

GKP Pauli-$X$ and $Z$ gates are implemented by position and momentum shifts, $\X\pqty{s} \coloneqq e^{-i s \p}$ and $\Z\pqty{t} \coloneqq e^{i t \q}$, for $s = \alpha$ and $t = \pi/\alpha$.  For GKP codewords, these shifts act as the intended Pauli gates on the logical qubit: $ \op{X}(\alpha)\ket{\psi_\text{GKP}} = \X_\logic \ket{\psi}[\logic] \otimes \gket{+_\text{GKP}} \label{eq:logicalXOnGKP}$, and $\op{Z}(\pi/\alpha) \ket{\psi_\text{GKP}} =\Z_\logic \ket{\psi}[\logic] \otimes \gket{+_\text{GKP}} $.  Arbitrary shifts act asymmetrically due to the choice to decompose in the $\op q$ basis.  Since $ \Z\pqty{t}$ is diagonal in $\op q$, momentum shifts have a tensor-product structure: $ \Z\pqty{t}  =  e^{i t \alpha/2 } \op{R}^z_{\logic} (t \alpha) \otimes e^{i t (2\alpha \mGop + \uGop)}$, where  $\hat{R}^z_\logic(\theta) \coloneqq e^{-i \theta \Z_\logic/2}$  performs a rotation by $\theta$ around the logical $z$-axis.  Arbitrary position shifts, $\op{X}(s)$, in general entangle the subsystems, a feature that will be explored in future work \cite{GiacMeni20}.  However, any position shift on a GKP state that changes the value of $\ell$ implements a logical $\op{X}_\logic$, with differences only in how the gauge mode is affected.  Although small momentum shifts have logical effects,  both position- and  momentum-shifted GKP states maintain their tensor-product structure.

The shift freedom of the GKP encoding is the foundation for its resistance to errors. Shift errors in position by less than $|\alpha/2|$ (and in momentum by less than $|\pi/2\alpha|$) are  perfectly correctible \cite{Glancy2006}.  The subsystem description based on modular position tells us why: small position shifts do not disturb the logical qubit
---they act solely on the gauge mode, as do their corrections. Small momentum shifts rotate the qubit by an amount revealed by measuring the gauge mode; displacing the full state back undoes this rotation.   The asymmetry in shift errors translates to biased noise at the logical-qubit level, which may be exploited for improved code performance \cite{WebsBartPoul15, Shru19, GuilMirr19, HangHeinKoen20}.   GKP error correction corresponds to measurement of the gauge mode in the modular representation of Ref.~\cite{ketterer2016} and returns it to $\ket{+_{\text{GKP}}}_G$ through displacements.

{\bf Approximate GKP states.---}The subsystem decomposition identifies the logical state in imperfect bosonic encodings, allowing quantitative characterization of the encoding quality using the logical fidelity. Finite-energy approximations to GKP states have nonorthogonal CV wavefunctions~\cite{gkp2001, Terhal2016, Motes:2017aa}; thus, it is unclear that such states faithfully encode intended qubit states.

Consider pure, approximate GKP states, whose position-basis wavefunction ${\psi}_\text{GKP}(s)$ is given by a $2\alpha$-periodic superposition of Gaussian spikes, each with variance $\Delta^2$. The complex amplitudes of every other spike are determined by the  intended  qubit state $\ket{\psi} = a\ket{0} + b \ket{1}$.  A comb of normalized Gaussians can be written as $\sum_{m \in \mathbb{Z} } G_{\Delta^2}(s-2\alpha m) = \frac{1}{2\alpha} \vartheta (\frac{s}{2\alpha}, \frac{ 2\pi i  \Delta^2}{4 \alpha^2} )$, where  $\vartheta\pqty{z,\tau} \coloneqq \sum_{m \in \integers} \exp\bigl[ 2\pi i \bigl(\tfrac 1 2 m^2 \tau + m z\bigr) \bigr]$  is a Jacobi theta function of the third kind. A broad Gaussian envelope with variance $\kappa^{-2}$ damps spikes far from the origin, making the state physical~\footnote{For $ \ket{\tilde{+}_\text{GKP} } $, this parameterization gives a momentum-space wavefunction with $\frac{2\pi}{\alpha}$-periodic spikes, spike variance $\kappa^2$, and envelope variance $\Delta^{-2}$.}.  We compactly express an approximate-GKP wavefunction as
	\begin{equation}
		\psi_\text{GKP}(s) = \frac{e^{- \frac{\kappa^2}{2} s^2}}{\sqrt{\mathcal N}} \left[ a \vartheta \Big(\frac{s}{2\alpha}, \tau_\Delta \Big) + b \vartheta \Big( \frac{s-\alpha}{2\alpha}, \tau_\Delta \Big) \right] ,
	\end{equation}
with $\tau_\Delta \coloneqq \frac{i \pi  \Delta^2}{2 \alpha^2}$ and normalization $\mathcal N$ \cite{Matsuura:2019aa}.  The limit $\Delta, \kappa \rightarrow 0$ gives a normalized, ideal GKP state~\footnote{In this limit, the ideal GKP states are normalized by an infinite constant.}.

%
\begin{figure}
    \includegraphics[width=0.45\textwidth]{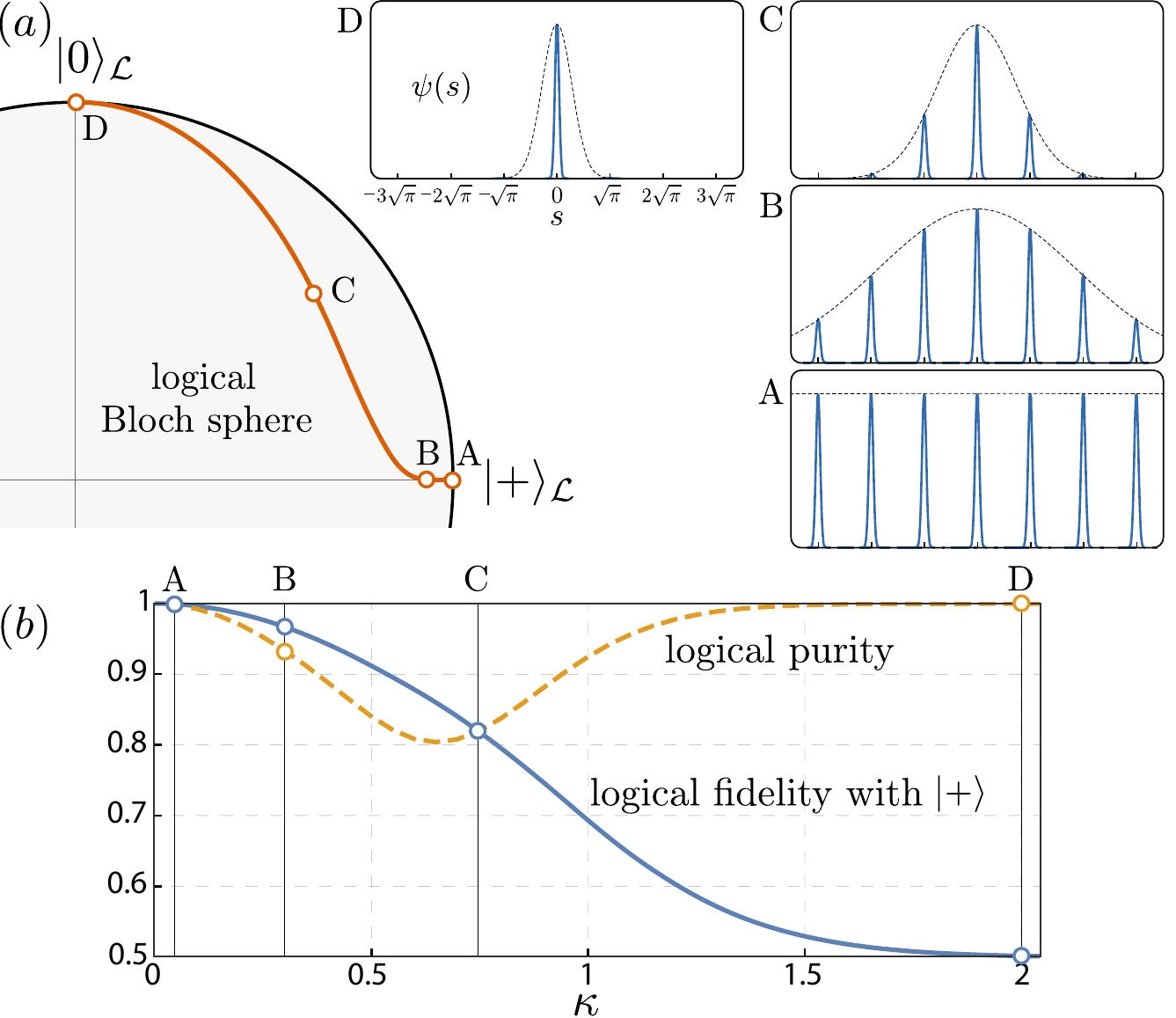}
    \caption{
    \label{fig:appgkp}
 Encoding quality of approximate square-lattice GKP states ($\alpha = \sqrt{\pi}$) intended to encode the qubit state $\ket{+}.$
 (a) Logical Bloch vector in the $xz$-plane for  $|+_\text{GKP} \rangle$ states with fixed spike width $\Delta = 0.1$ and varying envelope parameter $\kappa$.
Wavefunctions (solid blue) and envelopes (dashed black) are shown for $\kappa \in \{0.05, 0.3, 0.75, 2 \}$ to illustrate various approximate GKP states (A--D).
 For large envelopes, $\kappa \ll 1$, the logical qubit faithfully encodes $\ket{+}$.
For small envelopes, $\kappa \gg 1$, the logical state approaches $\ket{0}[\logic]$, because the envelope damps all spikes except the one centered at zero.  (b) Encoding quality captured by the {logical fidelity}, Eq.~(\ref{eq:logicalfidelity}), and logical purity $\Tr [ (\op{\rho}_\logic)^2]$.
}
\end{figure}

 Each approximate GKP state encodes a (potentially mixed) logical state $\op{\rho}_\logic$ whose encoding quality is determined by the logical fidelity with the intended qubit state, $\mathcal{F}_\logic (\op{\rho}_\logic, \ketbra{\psi}{\psi})$. Figure~\ref{fig:appgkp} investigates encoding quality for $| +_\text{GKP} \rangle$ states ($a=b=\frac{1}{\sqrt{2}}$).  For very good position spikes, $\Delta \ll \alpha$, the logical state is
\begin{eqnarray} \label{eq:smallspikelogicalstate}
   \op{\rho}_\logic  & =  \frac{1}{2 \vartheta\pqty{0,4\tau'}}
    \pmqty{
        \vartheta\pqty{0, \tau'}
        &
        e^{\frac{-\alpha^2 \gkpkappa}{4}}\vartheta\pqty{ \tfrac{1}{4}, \tau'}
        \\
       e^{\frac{-\alpha^2 \gkpkappa}{4}} \vartheta\pqty{ \tfrac{1}{4}, \tau'}
        &
        \vartheta\pqty{ \tfrac{1}{2}, \tau'}
    } \, ,
\end{eqnarray}
where $\tau' \coloneqq \frac{1}{2}\tau_\Delta(\Delta^2 + \kappa^{-2})$.  This state limits to $\ket{+}[\logic]$ for small~$\kappa$ (large envelope) and to $\ket{0}[\logic]$ for large~$\kappa$ (small envelope), in agreement with Fig.~\ref{fig:appgkp}(a).

From the perspective of the subsystem decomposition, approximate GKP states describe imperfectly encoded qubits in an ideal translation-symmetric code, just as GKP originally interpreted them \cite{gkp2001}.

{\bf Hidden qubit cluster states.---}Continuous-variable cluster states (CVCSs) enable universal, fault-tolerant quantum computing when used with GKP-encoded qubits \cite{Menicucci:2014cx}.   While quantum computing with CVCSs has no built-in definition of a qubit~\cite{Menicucci2006}, we have shown above that $\ket{0}_p$ momentum eigenstates, which comprise the CVCS, encode logical $\ket{+}$ states.  Here, we decompose the CV unitary gates between modes to reveal that a CVCS contains a logical-qubit cluster state entangled with the gauge modes.   Modular-position measurements \cite{Fluhmann:2018aa, Weigand:2019aa} disentangle the logical-qubit cluster state.

A CVCS is constructed by entangling modes pairwise using the two-mode gate $\CZ\bqty*{g} \coloneqq e^{i g \q \otimes \q}$. Decomposing this gate in the subsystem basis gives a product of nine commuting gates: $\CZ\bqty*{g} = \prod_{A,B} \CZgen{g} {A,B}$, where $\CZgen{g}{A,B} \coloneqq e^{i g \op{A} \otimes \op{B}}$ is a generalized controlled-$Z$ gate,
with labels $A,B$ each ranging over $\{ \alpha \ell, \uG, 2\alpha \mG \}$. For $g=1$ and $\alpha = \sqrt{\pi}$,  products giving  $2 \pi \times \text{integer}$  reduce to the identity, leaving
\begin{equation}
\CZ\bqty*{1}    \label{eq:cz}
	    =  \CZL \CZgen{1}{\{\sqrt{\pi}\ell,\uG\}}  \CZgen{1}{\uG,\uG}  \CZgen{2}{\{\sqrt{\pi}\mG,\uG\}} ,
\end{equation}
where $\CZL \coloneqq \exp\pqty*{i \pi \op{\ell} \otimes \op{\ell}}$ is a logical controlled-$Z$ gate, and $\CZgen{g}{\{A,B\}} \coloneqq \CZgen{g}{A,B} \CZgen{g}{B,A}$.  The second gate above entangles the logical subsystems with the gauge modes, and the two final gates act solely on the gauge modes.

\begin{figure}[t]
\includegraphics[width=0.46\textwidth]{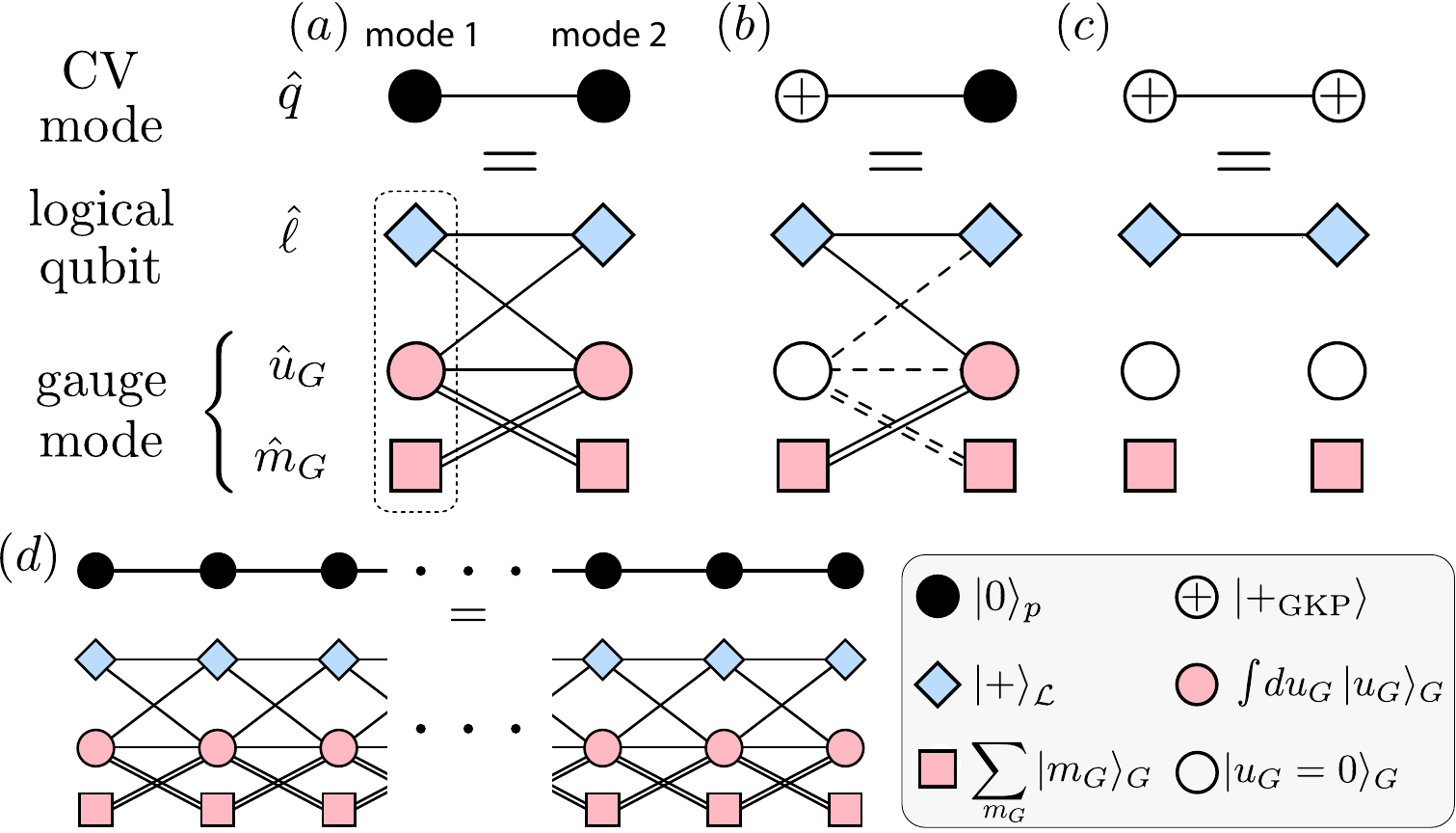}
\caption{
    \label{fig:CSdecomposition}
Graphical representations of states in the subsystem decomposition.
Different line styles indicate edge weights from Eq.~\eqref{eq:cz}.
(a) Logical-gauge entanglement structure for an infinitely squeezed two-mode squeezed state. The decomposition of a single-mode state $\pket{0}$ is grouped inside the dotted lines.
(b) A modular-position measurement on mode 1
disconnects the middle node, indicated by dashed lines. (c) Then, a similar measurement on mode 2 fully disentangles the logical two-qubit cluster state. (d) Entanglement structure in a linear CVCS.
}
\end{figure}

A pedagogical case is an ideal two-mode CVCS (locally equivalent to an EPR pair \cite{Einstein1935}), where
 each mode is initially prepared in a zero-momentum eigenstate, $\pket{0} = \ket{+}[\logic] \otimes \gpket{0}$ (see Table~\ref{tab:table1}). Decomposing the state using Eq.~\eqref{eq:cz} gives
	\begin{equation} \label{eq:hiddenqubitCS}
		\CZ\bqty{1} \pket{0}^{\otimes2} = \CZgen{1}{ \{ \sqrt{\pi}\ell,u_G \} } \big( \qubitCS{}[\logic] \otimes \gket{\Phi_2} \big) \, ,
	\end{equation}
 where ${\qubitCS{}[\logic] \coloneqq \CZL \ket{+}[\logic]^{\otimes 2}}$  is a ``hidden" two-qubit cluster state entangled with two-gauge-mode state, $\gket{\Phi_2} \coloneqq \CZgen{1}{\uG,\uG} \CZgen{2}{ \{ \sqrt{\pi}\mG,\uG \} } \gpket{0}^{\otimes 2}$.
The logical-gauge entanglement structure of this state is shown in Fig.~\ref{fig:CSdecomposition}(a). The top row depicts the two-mode CV state, and the lower rows give its subsystem decomposition, with gauge modes represented in the  partitioned-position basis (see Fig.~\ref{fig:CSdecomposition} legend). Edges represent generalized controlled-$Z$ gates with weight~1 (single line) and weight~2 (double line).

The logical qubits can be disentangled by resetting their modular position---similar to (partial) gauge-fixing \cite{Paetznick:2013aa,Brown:2016aa}. Regardless of the gauge-mode state, a modular-position measurement on mode 1 followed by a $u_G$-shift to $\uG = 0$ disconnects the middle node, Fig.~\ref{fig:CSdecomposition}(b),  and projects the CV mode into $\ket{+_\text{GKP}}$.    GKP error correction \cite{gkp2001} on just the $\op q$ quadrature realizes this measurement on a single mode; applied to both modes, it produces a GKP-encoded cluster state, which
features a disentangled logical-qubit cluster state [Fig.~\ref{fig:CSdecomposition}(c)].   Due to gauge freedom, there are other ways to disentangle the logical-qubit cluster state, for example by performing the inverse of the logical-gauge entangling gate followed by any gauge-mode unitary. However, this does not, in general, leave the CV modes in a GKP cluster state.

Decomposing a CVCS follows the same prescription.  An ideal, $N$-mode CVCS is given by ${\CVCS{\mat V} \coloneqq \CZgen{\mat{V}}{\vec{q},\vec{q}} \pket{0}^{\otimes N}}$, where  $ \CZgen{\mat{V}}{\vec{A},\vec{B}} \coloneqq \prod_{j<k} \CZgen{V_{jk}} {A_j,B_k}$ for length-$N$ vectors $\vec A$ and $\vec B$,  and $\mat V$ is an $N\times N$, real, symmetric, zero-diagonal adjacency matrix for a graph encoding the connections and interaction strengths~\cite{Menicucci2011}. When $\mat V$ has binary entries (an unweighted graph), we find a hidden, $N$-qubit cluster state $\qubitCS{\mat{V}}[\logic] \coloneqq \CZL [\mat{V}] \ket{+}[\logic]^{\otimes N}$ entangled with an $N$-mode gauge state $\gket{\Phi_N} \coloneqq \CZgen{\mat V}{\vec u_G, \vec u_G} \CZgen{2\mat V}{ \{ \sqrt{\pi} \vec m_G, \vec{u}_G \} } \gpket{0}^{\otimes 2}$  via the interaction operator $\CZgen{\mat{V}}{ \{\sqrt{\pi}\boldsymbol{\ell}, \vec{u}_G\} }$. Figure~\ref{fig:CSdecomposition}(d) shows a linear CVCS with $V_{i, i+1} = 1$. Modular-position measurements of the gauge modes (and subsequent $\uG$-corrections) disentangle the logical-qubit cluster state.

{\bf Conclusion.---}Continuous-variable quantum information is plagued by the curse of infinity---with \emph{ad hoc} methods designed to accommodate subspace-encoded qubits.  We have presented a broadly applicable framework---the subsystem decomposition---to divide a bosonic mode into virtual subsystems.  This technique is a flexible mathematical construction.
For example, the decomposition presented above can be performed on any quadrature using any bin size $\alpha$.  An entirely different subsystem decomposition based on binning Fock space can be used to analyze bosonic codes with discrete rotation symmetry~\cite{Grimsmo:2020aa, Baragiola:2019ab}.

Once a subsystem decomposition is performed, the gauge trace identifies the logical qubit, giving access to the tools of qubit quantum information \cite{TzitBourMeni20,Wan:2019aa}.  One can go further still: decompose any CV quantum operation to uncover its action on the subsystems. For bosonic error-correcting codes, this procedure gives a direct map from CV noise to logical error channels \cite{GiacMeni20}.  The framework also has applications beyond quantum computing; for example, to study non-Markovian reduced-state dynamics of the logical qubit.

While we focused here on qubits, the immense Hilbert space of a bosonic mode allows for other discrete subsystems. Encoding a logical qudit  in the paritioned-position basis  is straightforward: define the logical basis states using $m \text{ mod } d$, where $d$ is the qudit dimension. A different extension encodes multiple qubits in a ``Russian nesting doll'' fashion: decompose the gauge mode using the parity of $\mG$, and then repeat this process.

{\bf Acknowledgments}---%
We thank Lucas Mensen, Blayney Walshe, Kwok Ho Wan, Krishna Kumar Sabapathy, Joshua Combes, Arne Grimsmo, Rafael Alexander, Victor Albert, and Takaya Matsuura for discussions. This work was supported by the Australian Research Council Centre of Excellence for Quantum Computation and Communication Technology (Project No.\ CE170100012).

\bibliography{MenicucciPapersRefs.bib}
\bibliographystyle{apsrev4-2_title}

\end{document}